\documentclass{desyproc}
\newcommand{\pom}{$\, \mathrm{I \!\!\! P} \, $}

\newcommand{\met}{E$\!\!\!\!/_T$}
\usepackage{subfigure}
\usepackage{graphicx}
\usepackage{caption}
\usepackage{wrapfig}
\begin{document}
\title{Precision proton spectrometers for CMS}

\author{{\slshape Michael Albrow$^1$}\\[1ex]
$^1$Fermilab. Wilson Road, Batavia, IL 60510, USA\\
}



\acronym{EDS'09} 

\maketitle

\begin{abstract}
We plan to add high precision tracking- and timing-detectors at $z = \pm \sim 240$m to study central exclusive processes
$p + p \rightarrow p + X + p$ at high luminosity. This enables the LHC to be used as a tagged $\gamma\gamma$ collider,
with $X = \ell^+\ell^-, W^+W^-$ and as a ``tagged" gluon-gluon collider (with a spectator gluon) for QCD studies with jets.
A second stage with $z \sim$ 420m would allow observations of exclusive Higgs boson production.
\end{abstract}

 Adding forward proton spectrometers to CMS, with tracking precision $\sim 10 \; \mu$m and timing precision $\sim$10 ps, enhances the physics
 potential of CMS by enabling the study of specific clean exclusive reactions, both Standard Model and beyond. A proposal to do this, now
 called Precision Proton Spectrometers, PPS, (it was previously FP420~\cite{fp420}) has now been endorsed by the CMS Management and Collaboration
 Boards.  In $p + p \rightarrow p + X + p$, where ``+" means a large rapidity gap, the protons
 typically have $p_T \lesssim$ 1 GeV and $\xi = 1 - (p_z/p_{beam}) \lesssim$ 0.05, and they go down the beam pipes. 
 They are deflected out of the beam by LHC magnets, and at 210 m are a few mm to the side. Silicon tracking detectors and timing
 detectors will be installed there using special movable vacuum chambers, either a sideways-moving beam pipe with a thin flat wall, or a Roman
 pot.
 
 The four-momentum transfer from the protons can be carried only by a virtual photon, $\gamma$, or by a pomeron \pom .
 The pomeron is a feature of Regge theory, a strongly-interacting color-singlet, at leading order a pair of gluons. 
 High-$Q^2$ processes such as high-$E_T$ jet production can be calculated as perturbative $gg \rightarrow gg$ together with another gluon
 exchange to cancel the color and allow the protons to emerge intact. The non-perturbative QCD involved (low-$x$ gluon fluxes, Sudakov
 suppression of soft gluon radiation, rapidity gap survival probability) is tested by such measurements, and the same physics is needed
 to calculate exclusive H(125) production with $gg \rightarrow H$ through a top loop. Photon-pomeron interactions can produce exclusive Z-bosons
 through virtual $q$-loops. The two-photon process $\gamma\gamma\rightarrow W^+W^-$ is a window on BSM
 physics, being sensitive to triple- and quartic- gauge-boson couplings. Indeed CMS has observed two candididate events with $e^\pm \mu^\mp$ and
 missing $E_T$ and no other tracks, putting stringent limits on these couplings, but without detecting the protons. The PPS will enable much
 more data, and measuring the protons provides kinematic constraints and kills background.
 
 Pomeron exchange is expected to dominate when the scattered protons have $\xi \lesssim$ 0.05, corresponding to central masses 
 (with both protons detected): $M(X) \lesssim$ 3 GeV at the ISR
 (glueball searches), 100 GeV at the Tevatron (jets) and up to 700 GeV at the LHC(13 TeV). Thus for the first time the LHC gives access to electroweak-scale
 masses; interesting channels are $X = Z, H, WW, ZZ, \mu\mu$ and of course jets. With only one proton detected, high mass single diffraction can be studied,
 but probably only in low pile-up runs when one can associate the proton with the right central collision (out of perhaps 40!).
 
 The possibility of seeing exclusive Higgs production was first pointed out in Ref.\cite{alb1} and followed with a Letter of Intent \cite{loi}
 to the Fermilab PAC in 2000. At that time there were orders of magnitude differences in the cross section predictions and the LoI was not
 followed with a proposal; rather we made a  series of measurements in CDF of exclusive $\gamma\gamma$~\cite{cdfgg}, $\chi_c$~\cite{cdfchic} and 
 dijet production, which
 reduced the uncertainty on $\sigma(H)|_{exclusive}$ to a factor about $^\times_\div 3$. It is much too small for the Tevatron, but can be a few fb
 at the LHC. Annual meetings at Manchester led to the FP420 project~\cite{fp420}. After 2006 ATLAS continued to develop it as AFP and CMS
 as HPS (now PPS). 
 
 Traditionally Roman pots have been used (since the CERN-Rome ISR experiment in 1971) to place detectors (in air) very close to the beams.
 Another technique, a sideways-moving beam pipe, was used at HERA. Although there is less experience with an MBP, it has the 
 advantages of allowing long (e.g. 50 cm) pockets for easily accessible detectors, there are no differential pressure forces, and there is less effect (through impedance)
 on the beams. In addition, at 420 m the deflected protons are in-between the two beam pipes and only MBPs can fit. Both options are still being
 pursued, and will (hopefully) be tested with high luminosity in 2015. 

 The PPS has to reach small cross sections (fb) and therefore to run with high pile-up conditions, with 30-40 interactions per bunch crossing
 (every 25 ns). There will be a large pile-up background in which two single diffractive protons from different collisions in the same bunch crossing
 are detected. Kinematic constraints can be applied, e.g. the missing mass from the incoming and outgoing protons must be consistent with 
 the central mass, there should
 be no extra tracks $N(ass) = 0$ on the vertex (exactly two charged leptons for $W^+W^-$, no large $k_T$ tracks to jet axes, etc.), and the longitudinal momentum ($p_z$) difference
 between the protons should equal the $p_z$ of the central state. However a reduction factor $\sim \times 25$ can come from precision
 timing~\cite{alb1,loi}.
 If we can measure the \emph{time difference} $\Delta t$ between the protons with $\sigma(\Delta t) = \sqrt{2} \times \sigma_t = 15$ ps we 
 have $\sigma(z_{pp})$ = 2.25 mm, to be matched with $z_{vertex}$. We have developed quartz Cherenkov counters (\textsc{quartic})~\cite{quartic}
 that satisfy the requirements of being edgeless on the beam side, radiation hard, segmented in 24 3 mm $\times$ 3 mm cells, and covering the needed area
 (18 mm in $x$ and 12 mm in $y$).
 For the PPS we developed the L-bar \textsc{quartic}, with an array of 24 bars with SiPMs. We have demonstrated $\sigma_t$ = 30 ps in the
 test beam; the plan is to have four modules in line to achieve $\sigma_t$ = 15 ps, with improvements possible. A 24-channel module is being prepared for beam
 tests in November.
 
 To give an idea of the acceptance in $\xi,t$ of a small detector at $z$ = +240 m, Fig. 1a shows contours at several fixed $t$ and $\xi$ values in
 $x,y$. The inner edge is shown at 2 mm from the beam. At $\xi$ = 0.05 one has 100\% acceptance (all $\phi$)
 out to $t$ = 2 GeV$^2$. Coulomb-scattered protons (e.g. photoproduction of $Z, \gamma\gamma \rightarrow W^+W^-$) have very small $|t|$ and thus $y\sim 0$,
 which could be used in a trigger. At $z$ = +250 m the contours are slightly different, as angles $\theta_x, \theta_y$ are not zero, so we plan two stations
 about 10 m apart to give $\sim 1 \;\mu$rad angular precision. Fig.1b shows the approximate acceptance, 
 assuming 3 mm from the beam, in $M(X)$ for small $|t|$ for three cases:
 detector stations at $z$ = 240 m on both sides (Stage 1), and together with additional stations at 420 m (Stage 2). Stage 1 is good for $W^+W^-$, other
 massive states and high $E_T$ jet physics, while 420 m stations are needed to see H(125) with both protons. 
 The configuration 420+240 is best for acceptance, while 420+420 has the
 best mass resolution because of the 120 m of 8T dipoles between 240 m and 420 m.
 In Stage 1 we will study both QCD and EWK physics.

\begin{figure}[ht]
\centering

\subfigure{%
\includegraphics[width=60mm]{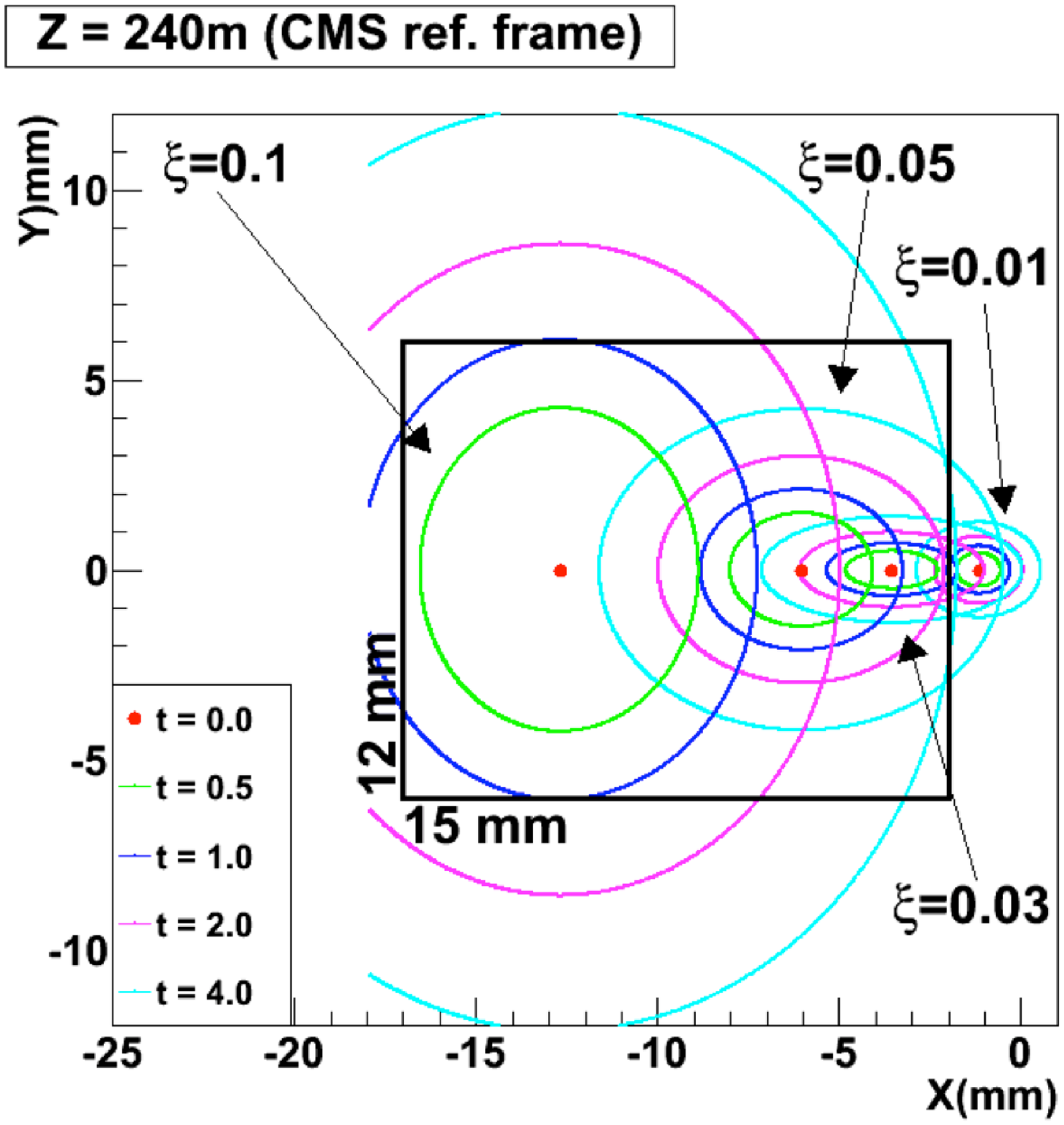}}
\label{fig:txi}
\quad
\subfigure{%
\includegraphics[width=60mm]{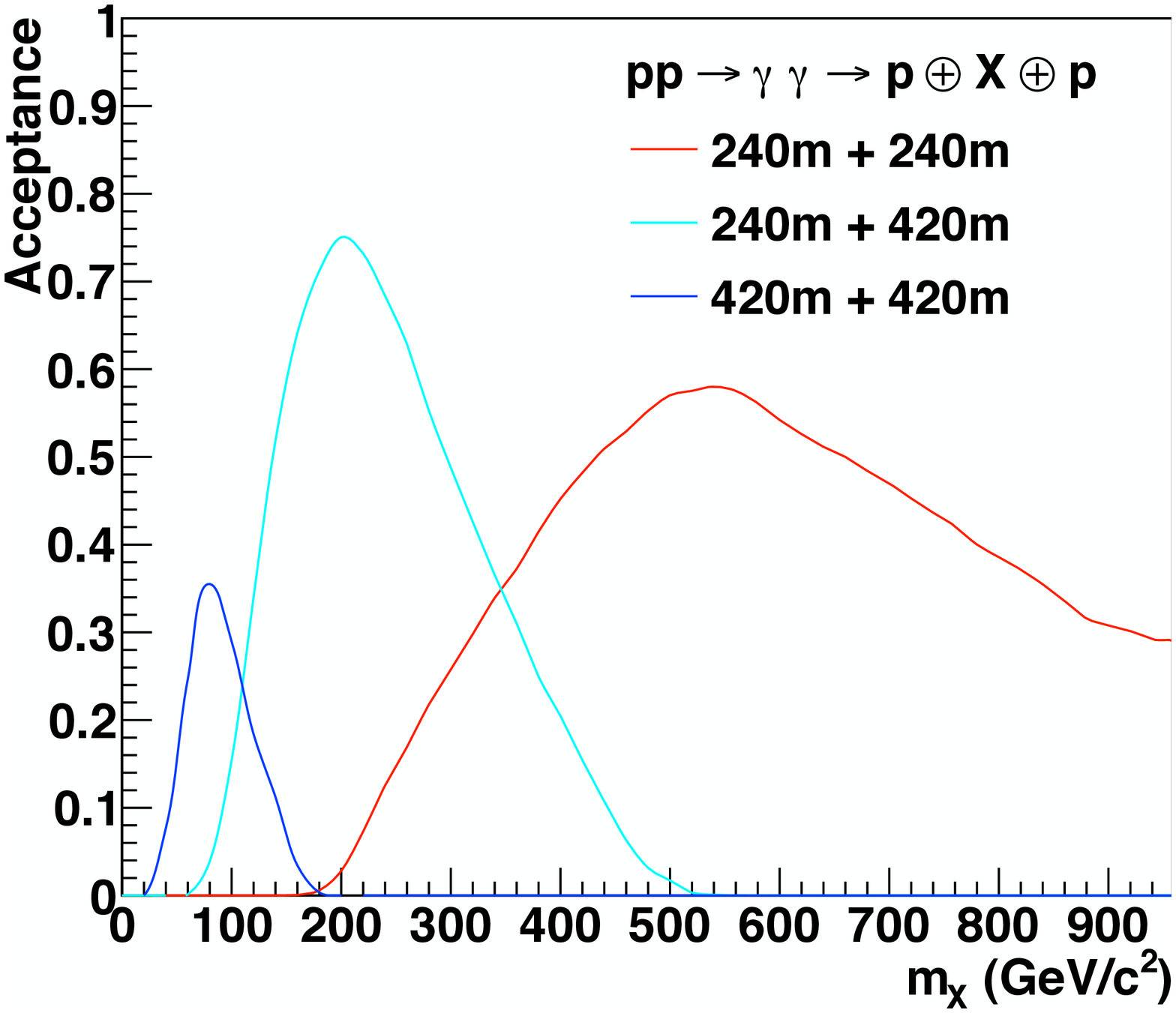}}
\vspace{-0.4in}
\caption{(Left) Contours at fixed $t$ and $\xi$ in the $x,y$-plane at $z$ = 240 m, 
low-$\beta$. (Right) Acceptance in $M(X)$ for $p + X + p$ with Stage 1 detectors. It is shown for 
two-photon interactions (very small $|t|$).}
\label{fig:massacc}

\end{figure}

 QCD: Exclusive di-jets and tri-jets with $M(JJ,JJJ)$ up to $\sim 800$ GeV can be studied.
 The dijets should be pure gluon jets (the LHC as a ``gluon jet factory"), with a small component of $b\bar{b}$. 
 Both jets can be $b$-tagged; this is an important test of the $J_z = 0$ rule
 which states that exclusive $q\bar{q}$-dijets are forbidden for massless quarks. The rule is important in keeping the background to exclusive $H \rightarrow
 b\bar{b}$ small. The exclusive $q\bar{q}$-dijet spectrum is important as it constitutes the irreducible background to $H \rightarrow b\bar{b}$, as well as
 being a good test of the QCD mechanisms involved. Exclusive tri-jets can be $ggg$ or $gq\bar{q}$, distinguishable (statistically) by kinematics (the former
 are more Mercedes-like), and democratic between the five light-quark flavours. A foretaste is shown in Fig.2, one of a few dozen events with
 both protons detected in TOTEM Roman pots and two jets with $E_T > 20$ GeV in CMS, and very little else, from a short high-$\beta^*$ run in 2012. One week of dedicated low pile-up running could
 increase that data sample by a factor $>$1000! Most QCD events will not be ``golden" exclusive jets, but high mass \pom + \pom $\rightarrow$ hadrons,
 which is unexplored territory, not to be neglected. Study event shapes (spherical, very high multiplicity or ...), double parton scattering, heavy flavour
 production, etc. Developing event generators. like \textsc{pythia} but for $X$ in $p + X + p$ events is a challenge for theory and phenomenology and should lead to a
 better understanding of the strong interaction.

 EWK: Searches for central exclusive production of MSSM Higgs are on the agenda~\cite{tasevsky}. It is also very interesting to measure $\gamma\gamma \rightarrow W^+W^-$. 
 This uses the LHC as a photon-photon collider, with reach up to about 1 TeV, much higher than LEP and unique for the forseeable future.
 Two $W^+W^-$ candidates have been seen by CMS in the $e+\mu$+\met  channel without the protons, with no additional tracks. With proton tagging other $W$-decay
 channels can be used, proton-dissociation can be excluded, and a few hundred events obtained. The two observed events already put much more stringent limits
 on anomalous quartic gauge couplings than previous experiments (LEP); this is a unique study. What about exclusive $ZZ$ production, negligible in the Standard
 Model? Any exotic heavy states with vacuum quantum numbers (and even spin) that couple even indirectly (like $H$) to gluons may be visible.
 Even in Stage 1 there is some acceptance for $H(125)$ if only one proton is accepted (and the Higgs is boosted). But without both protons we do not have the
 kinematic constraints or $\Delta t_{pp}$ to kill background. The background is very small in $H \rightarrow WW^*$ (leptonic decays) and 
 $H \rightarrow \tau^+\tau^-$ but unfortunately the signals are tiny. So exclusive $H(125)$ studies need Stage 2. It seem obvious that we should study this
 particle every way we can, which should be enough justification for the modest outlay, even if its mass, spin and other properties will already be 
 established. However its parity, which must be $P = +1$, may not have been measured without assuming CP-conservation in the Higgs sector.
 Even seeing it exclusively demonstrates $P = +1$ unambiguously. (If it has mixed parity that would be \emph{very} interesting! 
 The direct coupling $Hb\bar{b}$ is measurable in a clean way. Stage 1 can also measure $Z$-photoproduction in $e^+e^-, \mu^+\mu^-$ and $\tau^+ \tau^-$,
 together with the continuum in these channels from $\gamma\gamma$. No surprises are expected, but these channels are good
 ``tools" for calibrating the momentum scales of the PPS trackers, as both proton momenta are predicted from the $\mu^+\mu^-$.
 
 To conclude, the addition of small but very precise tracking and timing detectors far along the beams, PPS at CMS, opens up a whole
 new field in both QCD and electroweak physics for a very modest cost. We have developed suitable detectors, state-of-the-art but very small (cm$^2$) to cope
 with pile-up around 30 interactions per crossing, more for $W^+W^-$.

\begin{figure}
\centering
\includegraphics[width=80mm]{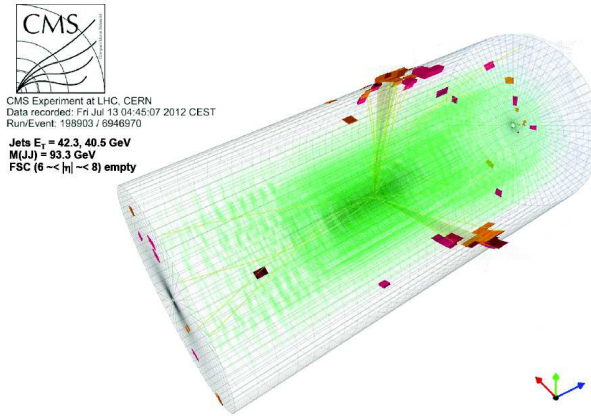}
\caption{Dijet event recorded by TOTEM and CMS in low pile-up (90 m $\beta^*$) run at $\sqrt{s}$ = 8 TeV.
Two leading protons and two jets with $E_T >$ 20 GeV were required.}
\label{fig:dijet}
\end{figure}

I acknowledge funding from the U.S. Dept. of Energy, and many people contributing to the development of the PPS,
and theoretical colleagues, especially Valery Khoze and Misha Ryskin.

\begin{footnotesize}

\end{footnotesize}
\end{document}